\documentclass[conference]{IEEEtran}
\IEEEoverridecommandlockouts
\usepackage{cite}
\usepackage{amsmath,amssymb,amsfonts}
\usepackage{graphicx}
\usepackage{textcomp}
\usepackage{xcolor}
\usepackage{hyperref}
\usepackage{relsize}

\def\BibTeX{{\rm B\kern-.05em{\sc i\kern-.025em b}\kern-.08em
    T\kern-.1667em\lower.7ex\hbox{E}\kern-.125emX}}

\definecolor{new}{rgb}{0.0, 0.5, 0.3}

\usepackage{algorithm}
\usepackage{algpseudocode}
\algrenewcommand\textproc{} 

\usepackage{fancyvrb}
\definecolor{cb_blue}{RGB}{7, 129, 230}
\definecolor{cb_green}{RGB}{9, 224, 186}
\definecolor{cb_lime}{RGB}{100, 212, 42}
\definecolor{cb_highlight}{RGB}{252, 202, 131}

\newcommand{\reducedstrut}{\vrule width 0pt height .9\ht\strutbox depth .9\dp\strutbox\relax}
\newcommand{\highlight}[1]{\begingroup
	\setlength{\fboxsep}{0pt}\colorbox{cb_highlight!50}{\reducedstrut#1\/}\endgroup
}

\begin{document}

\title{
SNARKChain: Proof-of-Useful-Work Blockchain Consensus with General-Purpose SNARK Marketplace
}

\author{
\IEEEauthorblockN{Samuel Oleksak, Richard Gazdik, Martin Pere\v{s}\'{i}ni, Ivan Homoliak}
\IEEEauthorblockA{
\textit{Brno University of Technology}\\
\textit{Faculty of Information Technology}\\
Brno, Czechia}
}

\maketitle

\begin{abstract}
	Proof of Work (PoW) is widely regarded as the most secure permissionless blockchain consensus protocol.
	However, its reliance on computationally intensive yet externally useless puzzles results in excessive electric energy wasting.
	To alleviate this, Proof of Useful Work (PoUW) has been explored as an alternative to secure blockchain platforms while also producing real-world value. Despite this promise, existing PoUW proposals often fail to embed the integrity of the chain and the identity of the miner into the puzzle solutions, not meeting the necessary requirements for PoW and thus rendering them vulnerable. In this work, we propose a PoUW consensus protocol that computes client-outsourced SNARK proofs as a byproduct, which are simultaneously used to secure the consensus protocol.
	We further leverage this mechanism to design a decentralized marketplace for outsourcing SNARK proof generation, which is, to the best of our knowledge, the first such marketplace operating at the consensus layer while meeting all necessary properties of PoW.
\end{abstract}

\section{Introduction}\label{sec:intro}
Permissionless blockchain systems rely on consensus protocols to provide agreement among mutually distrusting nodes.
The first and still one of the most popular blockchain consensus protocols is Proof of Work (PoW)~\cite{bitcoin}.
PoW secures the network from Sybil attacks by requiring the nodes that participate in consensus to solve computationally intensive puzzles, such as finding a particular hash~\cite{bitcoin}, solving memory-hard problems~\cite{scrypt}, or completing randomized computational tasks~\cite{primecoin}.
The results of these puzzles typically do not have any use outside the consensus protocol, which leads to this protocol being labeled as wasteful and environmentally harmful.
Some reports~\cite{carbon-footprint} suggest that Bitcoin alone may be responsible for 65.4 megatonnes of CO\textsubscript{2} emitted per year, which is comparable to the country-level emissions of Greece.

To justify the energy consumption of PoW, the community has proposed puzzles that deliver value beyond securing the consensus protocol, with notable examples including Cunningham chains~\cite{primecoin}, Cuckoo cycles~\cite{aeternity}, DNA sequence alignment~\cite{coinami}, and deep learning model training~\cite{coinai}.
Protocols incorporating such useful computational puzzles are known as Proof of Useful Work (PoUW) protocols.
These protocols have not yet gained significant traction in the blockchain space, mainly due to the challenges in designing meaningful tasks that meet the core PoW requirements, such as the verifiability of solutions, adjustable hardness, solution replay attack resistance (freshness), and chain integrity embedding~\cite{pouw,useless-work}.

Zero-knowledge Succinct Non-interactive Arguments of Knowledge  (zk-SNARKs)~\cite{snark} are a powerful cryptographic tool that enables verifiable computation and privacy preservation by allowing a prover to convince a verifier of the truth of a statement with no interaction and without revealing the underlying data.

However, their usability is hindered by the fact that generating zk-SNARKs is both computationally and memory intensive, often requiring tens of gigabytes of RAM and taking tens of minutes even on high-end hardware, making them impractical to compute on resource-constrained devices.
To address this, various off-chain proof generation frameworks and marketplaces have been proposed, allowing users to delegate proof generation to external providers.
These solutions can be broadly grouped as (a) \textbf{centralized}~\cite{centralized-snark} and \textbf{decentralized}~\cite{nil-proof-market,risc-zero-bonsai}, depending on whether a trusted operator manages task allocation and payments or whether matching is achieved through a peer-to-peer protocol, and (b) those that operate at the \textbf{application layer}~\cite{mina-snarketplace}, where proof outsourcing is an optional service, or those that are embedded at the \textbf{consensus layer}, where proof generation is inseparable from maintaining the blockchain’s security.

In this paper, we tackle both the challenges of PoUW task design and zk-SNARK generation by proposing a novel consensus protocol where zk-SNARK generation constitutes the useful work, inheriting the security of PoW while producing useful cryptographic proofs as a byproduct.

To the best of our knowledge, no existing decentralized zk-SNARK marketplaces generate \textbf{general-purpose} zk-SNARKs at the \textbf{consensus layer}. Our proposed solution represents the first such system. While some protocols~\cite{mina} utilize zk-SNARK at the consensus layer, their purpose is different, and they are limited to a \textbf{single specialized circuit}.

\paragraph*{\textbf{Contributions}}
Our contributions are as follows:
\begin{itemize}
    \item[$\bullet$] We propose a novel approach to offload resource-intensive SNARK proof computation onto nodes using a blockchain based on the PoUW consensus protocol, creating the first decentralized marketplace for SNARK computation that provides security to the consensus protocol.
In particular, we address the challenge of delegating zk-SNARK proof computation involving private inputs in a public permissionless blockchain setting.
    \item[$\bullet$] We analyze and demonstrate the security properties of our proposed consensus protocol. \end{itemize}

\section{Background}\label{sec:background}
In this section, we present the necessary preliminaries to facilitate an understanding of the remainder of the paper, with a particular focus on zero-knowledge proofs. 

\paragraph{\textbf{Proof of Useful Work (PoUW)}}
This concept emerged to address the profound inefficiency of traditional Proof of Work consensus, where immense computational power is expended on puzzles with no intrinsic value outside of securing the blockchain.
The core idea of PoUW is to replace arbitrary puzzles with computational tasks that produce valuable results, such as complex scientific simulations~\cite{coinami} or algorithmic challenges, e.g., $k$-orthogonal vectors problem~\cite{pouw}. PoUW covers a class of algorithms that decrease the energy wasted while still reaping the security benefits of PoW. 

\paragraph{\textbf{Zero-Knowledge Proofs (ZKPs)}}
Zero-knowledge proofs~\cite{zkp-original} are a cryptographic system that enables one party (the prover) to convey the correctness of a statement to another party (the verifier) without disclosing any additional information about the statement.
Some ZKP systems also possess the property of \textit{succinctness}, whereby both the proof size and verification time are significantly smaller than those of the original statement.
This is especially useful in the blockchain environment, where proof storage and verification are replicated among all consensus nodes. 

ZKPs have a wide range of applications across blockchain and cryptography, including cross-chain interoperability~\cite{zkbridge,zkrelay}, lightweight blockchains~\cite{mina}, transaction mixing~\cite{tornado-cash}, L2 chain facilitation~\cite{zk-rollup}, and federated learning~\cite{federated-learning,federated-learning-2}.

\paragraph{\textbf{Zk-SNARK}}
Zk-SNARK (Zero-Knowledge Succinct Non-interactive Argument of Knowledge)~\cite{snark} is a type of zero-knowledge proof system enabling the generation of proofs that are significantly shorter than the original statements, while still allowing for efficient verification time.

The process of obtaining zk-SNARKs typically starts with a program in a high-level domain-specific language (DSL), such as Circom~\cite{circom} or Zokrates~\cite{zokrates}.
Arithmetic circuits are used to represent computations using mathematical operations, specifically addition and multiplication.
R1CS (Rank-1 Constraint System) is an intermediate format that represents the program in a system of linear equations that capture the relationships between inputs, intermediate values, and outputs.
The circuit takes two types of inputs: \textit{public inputs}, which are known to both the prover and verifier, and \textit{private inputs}, which are known only to the prover.
These private inputs are used to demonstrate knowledge of a secret without revealing it.
The \textit{witness} is the assignment of values to all private inputs and intermediate variables that satisfy the circuit constraints for a given public input.

One of the drawbacks of SNARKs is the need for a \textit{trusted setup}, which introduces a potential security risk.
This setup involves generating a pair of cryptographic keys used for proof generation and verification. During this process, intermediate data known as \textit{toxic waste} are created, and if not securely discarded, this data can be used to produce false proofs.
Nevertheless, this problem can be alleviated by \textit{secure multi-party computation} (SMPC), where multiple participants collaboratively contribute randomness and share responsibility for the setup.
However, SMPC comes at the cost of increased coordination complexity and longer setup times.

Generating a zk-SNARK proof is computationally expensive and often impractical for resource-constrained devices. On the other hand, verification is fast and is often performed on limited-capability devices or even on-chain.
To address this imbalance, proof generation is often offloaded to specialized nodes with powerful hardware.
To facilitate this outsourcing, proof generation marketplaces have emerged~\cite{nil-proof-market,risc-zero-bonsai}, enabling users to request specific proofs to be generated in exchange for fees.
 
\section{Problem Definition}\label{sec:problem}
\label{sec:pow-conditions}

Developing a feasible PoUW protocol has the potential to significantly enhance the energy efficiency of PoW systems without compromising their security guarantees.
Although several PoUW schemes have been proposed~\cite{pouw,coinai,ofelimos}, none have yet reached a level of maturity and security for real-world deployment.
The goal of our work is to design a PoUW protocol that addresses the challenges of previous designs and meets the required properties of PoW.

\paragraph*{\textbf{Required Properties of PoW}}
For any PoW protocol to be viable in an adversarial environment, its puzzle must retain critical properties~\cite{puzzle-properties} which must also hold for PoUW:

\begin{itemize}
    \item[$\bullet$] \textbf{Asymmetry}: 
    Due to mutual distrust between the nodes, each node has to verify the correctness of a puzzle solution itself.
    To achieve reasonable scalability, the verification of a solution must be very quick while finding it should take significant effort and time.

    \item[$\bullet$] \textbf{Granularity and Parametrization}: 
    The difficulty of a computational puzzle should be dynamically adjustable to achieve desired network parameters, e.g., block time.

    \item[$\bullet$] \textbf{Amortization-freeness}: 
    Producing multiple puzzle solutions should be proportionally more difficult than producing a single solution.

    \item[$\bullet$] \textbf{Independence}: 
    Solving one instance of the puzzle provides no computational advantage toward solving other instances.

    \item[$\bullet$] \textbf{Efficiency}: 
    Additional overhead of operating PoW protocol in terms of memory and network bandwidth resources should be low.

    \item[$\bullet$] \textbf{Unforgeability}: 
    A malicious prover should not be able to construct such puzzle tasks that allow him to precompute the solution.
    
    \item[$\bullet$] \textbf{Freshness}: 
    The puzzle scheme is resistant to replay attacks, which would enable reuse of existing solutions or stealing of recently computed solutions.

    \item[$\bullet$] \textbf{Deterministic}: 
    The cost of finding a solution should either be deterministic or have a predictable expected value.

    \item[$\bullet$] \textbf{Progress-free}: 
    The probability of finding a solution should be independent of the effort already spent on solving the puzzle.

    \item[$\bullet$] \textbf{Publicly Verifiable}: 
    The puzzle solution can be verified by any party without the participation of the prover (non-interactively).
\end{itemize}

\section{Proposed Approach}\label{sec:proposal}
\begin{figure}[t]
    \centering
    \includegraphics[width=\linewidth]{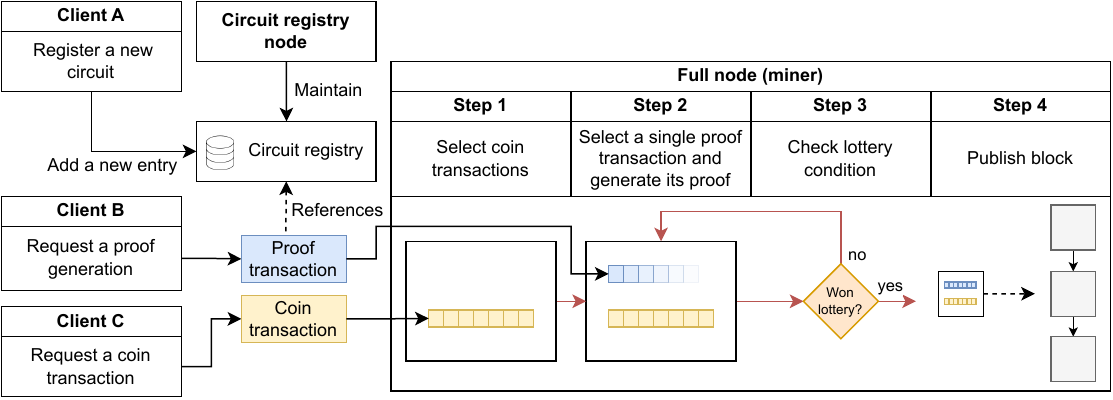}
    \caption{Diagram illustrating the lifecycle of a single block, including the different roles of users within the protocol.}
    \label{fig:block-lifecycle}
\end{figure}

In this section, we propose a novel PoUW consensus protocol in which clients request outsourced SNARK proof (i.e., without zero-knowledge property, since we only support public parameters) generation by providing an arithmetic circuit and proof parameters (see \autoref{fig:block-lifecycle}). Consensus nodes compute these proofs, establishing a decentralized general-purpose SNARK marketplace that combines proof generation with consensus.
The computation of these SNARK proofs forms the core puzzle of the consensus protocol, which secures the network. 
Each block within the chain contains two types of transactions: (1) \textbf{proof transactions}, which encapsulate SNARK generation and (2) \textbf{coin transactions} involving native currency operations.

\subsection{Network Participants}

In the proposed blockchain network, we distinguish three roles of participants: (1) \textbf{clients}, (2) \textbf{miners} and (3) \textbf{circuit registry nodes}.

\textit{Clients} do not participate in the network's consensus process, but they create and broadcast coin transfer transactions or utilize the marketplace properties of the network to request SNARK proof generation by publishing proof transactions.
Coin transaction fees are fixed (adjusted by the network), while proof transaction fees scale with the complexity of the associated arithmetic circuit.

\textit{Miners} (also referred to as \textit{full nodes} or \textit{workers}) participate in the consensus~-- they select coin and proof transactions from the corresponding mempools, generate the required SNARK proofs, publish new blocks, and validate incoming blocks.
Miners, in turn, receive a block reward and transaction fees by producing a block that contains the generated SNARK proofs and confirmed coin transactions.

The \textit{circuit registry nodes} maintain the SNARK circuit registry, which enables decentralized registration and retrieval of arithmetic circuits and their metadata.
They participate in SMPC for trusted setup during the initial circuit registration and store proving and verification keys.
To ensure reliability, these nodes are required to stake native currency, which can be slashed for misbehavior or inactivity (see \autoref{sec:circuit-management}).

\subsection{Block Proposer Selection}

The most important part of our approach is the selection of the block proposers, which is principally based on the idea of a weighted lottery, as in typical PoW consensus protocols.
In the long term, the frequency with which a block leader is selected should be proportional to their computational power (fairness property).
Nevertheless, some level of randomness must be incorporated into the system to prevent the most powerful miner from consistently winning each round by reaching the difficulty target first.
We begin by presenting a strawman approach, which is then incrementally refined.

\subsubsection{\textbf{Strawman Approach}}
Let us consider a system where (1) each coin and proof transaction can be included in only one valid block, (2) each computed proof is bound to the currently generated block by embedding block hash as a public parameter of a proof (see \autoref{sec:auth}), (3) the currently generated block is bound to the previous block through direct hash reference forming a linear chain, and (4) miners must generate enough proofs to reach a minimum difficulty threshold $\kappa$ for producing a block.
The value of $\kappa$ is periodically adjusted after a predetermined number of blocks (reflecting a certain amount of time, such as 2 weeks in Bitcoin) to account for changes in total mining power, maintaining an approximately constant expected block time.

\textbf{\textit{Shortcomings:}}
The approach misses the essential condition of progress-freeness, which mandates that the probability of successfully mining a block remains independent of the time already spent mining it.
The absence of this property would lead to block production being centralized, as each block would consistently be created by the strongest miner, who always reaches the threshold first, causing all other miners to have to repeatedly start over.
To improve it, we can introduce randomness into the block-building process.

\begin{figure*}[t]
    \centering
    \includegraphics[width=0.8\linewidth]{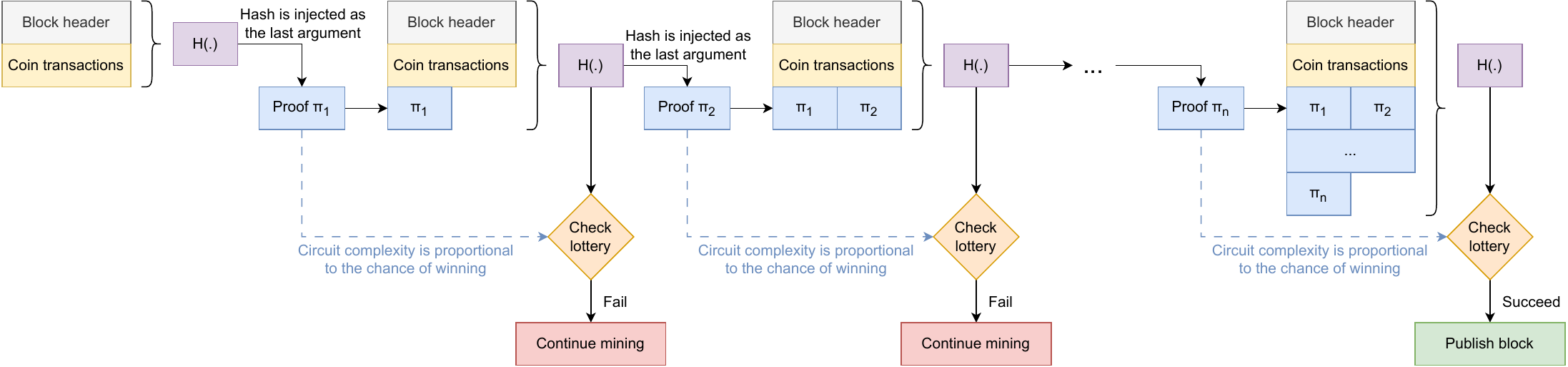}
    \caption{The lottery mechanism during block mining (as described in \autoref{sec:lottery}). A lottery is performed every time a new proof is added to the block in progress.} 

\label{fig:lottery}
\end{figure*}

\subsubsection{\textbf{Addition of Lottery}}
\label{sec:lottery}
To address the missing progress-freeness property of the Strawman, we replace condition (4) with a new condition (5) in which a lottery is triggered each time a miner adds a new proof to the current block being assembled, as illustrated in \autoref{fig:lottery}.
Due to the accumulative nature of a block in progress, each generated proof must be cryptographically bound to the previous proof within the same block, forming a \textit{proof chain} that prevents the theft of useful work solutions and ensures the integrity of the chain (see \autoref{sec:auth} for a more detailed description).
If a miner wins a lottery, the block is finalized and can be published.
The difficulty threshold for a single block is no longer a fixed value that must be reached every time, but rather an expected value in the long term, reflecting the mean effort required for block creation. 
This mechanism is designed to give smaller miners a fair chance to produce blocks while ensuring that, over time, the probability of block production aligns with each miner's proportional share of total mining power, under the assumptions of Poisson approximation, based on the law of rare events~\cite{laws-of-small-numbers} (see $\mathbf{H_2}$ in \autoref{sec:eval}).
The greater the complexity $C_i$ of the $i$-th proof $\pi_i$ within the proof chain, the higher the probability of winning the lottery $P_{win}$, as described by the following formula:
\begin{eqnarray}    
P_{win}(i) = \frac{C_i}{\kappa},
\end{eqnarray}
where $\kappa$ denotes the current block difficulty target, which is constant across the current block round.
This satisfies the progress-free property of PoW protocols to a certain degree since the miner’s chance of winning the lottery depends solely on the most recently generated proof.\footnote{The process of generating a single proof involves incremental progress, but it is reset once the proof is completed and the lottery is executed.}
Consequently, the time between block creations follows an exponential distribution (we elaborate on the expected block time in \autoref{sec:time-to-generate}).
The lottery is executed by comparing the hash of the potential block (as shown in \autoref{fig:lottery}) with a target threshold derived from $P_{win}$ similarly to the Bitcoin protocol.

It must be ensured that the computational complexity of each arithmetic circuit is proportional to its probability of winning the lottery, calibrated so that the expected difficulty remains consistent in the long term.
We define the complexity $C_i$ of a single proof as the number of gates (constraints) of the circuit associated with the proof transactions (see \autoref{sec:time-to-generate}).

\textbf{\textit{Shortcomings:}}
This approach produces a fair amount of useful work without giving an undue advantage to more powerful miners, ensuring that the relationship between work and reward retains the fairness property (see $\mathbf{H_1}$ in \autoref{sec:eval}).
However, most of the work performed would be wasted, as miners who do not win the round must discard their partial block along with the generated proofs in accordance with conditions (2) and (3) defined in Strawman.

\subsubsection{\textbf{Reducing Wasted Work -- Adjusting Lottery Odds}}
\label{sec:psi-parameter}
To slightly decrease the amount of wasted work, one could implement a slight advantage in lottery odds for miners that have performed more useful work (constructed a proof chain with more accumulated work) within the current block round.
This can be achieved by introducing an additional term into the lottery winning function that accounts for the sum of the complexities of all generated proofs in the currently produced block, where the influence of this term can be adjusted using the constant~$\psi \in [0; 1]$:
\begin{eqnarray}   
P_{win}(i) = \frac{C_i}{\kappa} + \psi \cdot \mathlarger{\sum}_{j < i} \frac{C_j}{\kappa}.
\end{eqnarray}
\indent
\textit{\textbf{Shortcomings:}}
Intuitively, this alteration provides a disproportionate advantage to the stronger miners who can produce stronger chains within a single block window, thereby increasing centralization (see $\mathbf{H_3}$ in \autoref{sec:eval}).

\subsubsection{\textbf{Reducing Wasted Work -- Naive Parallel Block Production}}

\begin{figure}[b]
    \centering
    \includegraphics[width=0.9\linewidth]{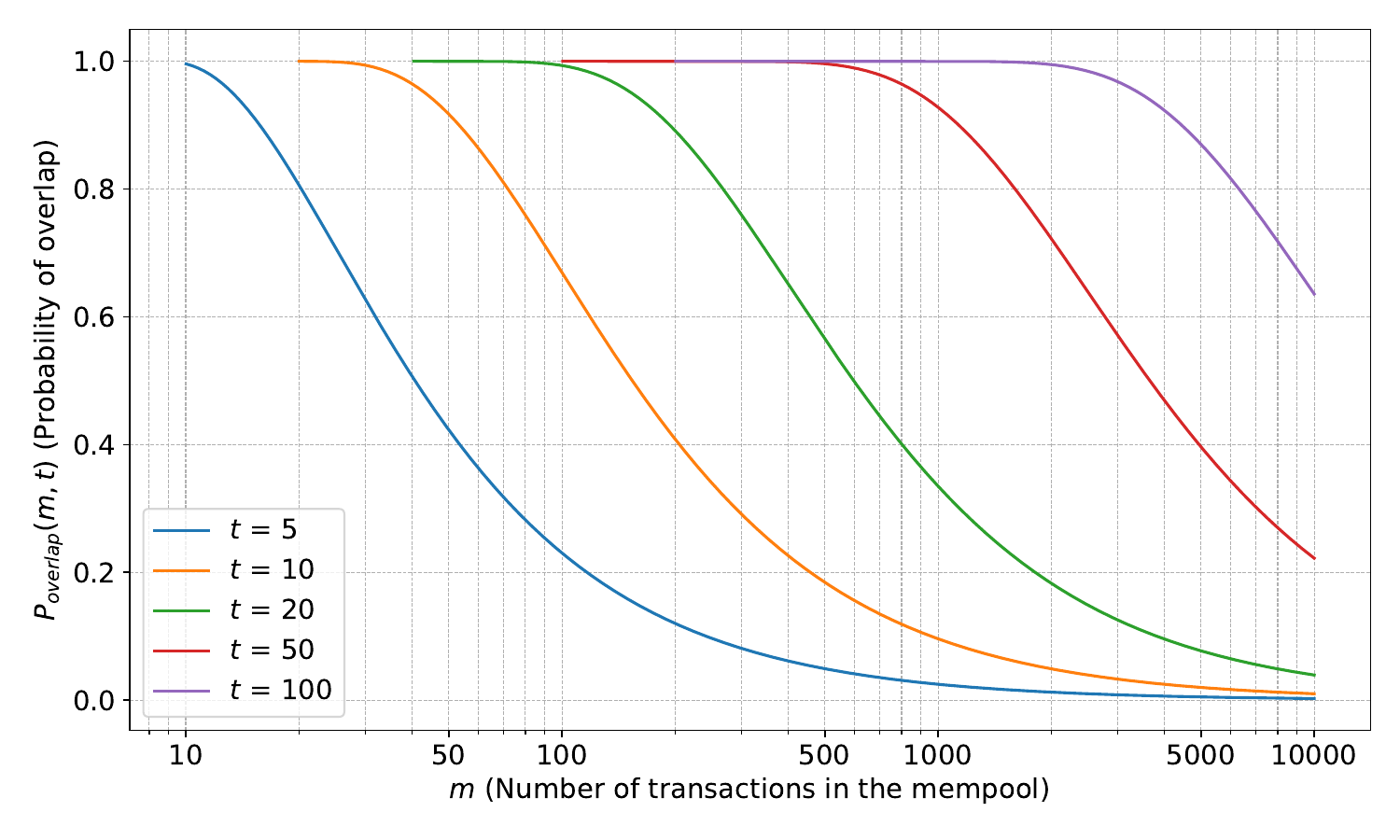}
    \caption{
        The overlap probability $P_{overlap}(m,t)$ of two randomly selected transaction sets of size $t$ from the mempool of size $m$9    }
    \label{fig:mempool-overlap-probability}
\end{figure}
By relaxing condition (3) of Strawman, which requires each newly generated block to be strictly bound to its immediate predecessor, forming a linear chain, while still maintaining condition (1) that each proof and coin transaction can be included only once, we arrive at a scenario where parallel block production becomes possible if overlaps are avoided.
If two miners were to concurrently produce a block of $t$ transactions\footnote{For simplicity, we assume that each proof transaction is equally difficult to compute and that transactions are selected at random.} with $m$ unconfirmed transactions in the network (where $m \geq 2t$), then the probability $P_{overlap}$ of two miners selecting subsets of transactions with at least a single transaction overlap would be:
\begin{equation}
P_{overlap}(m, t) = 1 - \frac{\binom{m-t}{t}}{\binom{m}{t}}.
\end{equation}
\indent
\textit{{\textbf{Shortcomings:}}}
Despite naively reducing wasted work by this approach, it still suffers from a large amount of wasted work caused by frequent overlaps of transactions, unless $m \gg t$, as can be seen in \autoref{fig:mempool-overlap-probability}.
The probability of overlap decreases as the total number of transactions in the mempool grows and increases with a higher number of transactions per block.
However, even with a large mempool and small blocks, the overlap probability remains significant, resulting in a substantial amount of wasted work.

\subsubsection{\textbf{Addition of Bucketing}}
\label{sec:bucketing}
\begin{figure}[t]
    \centering
    \includegraphics[width=\linewidth]{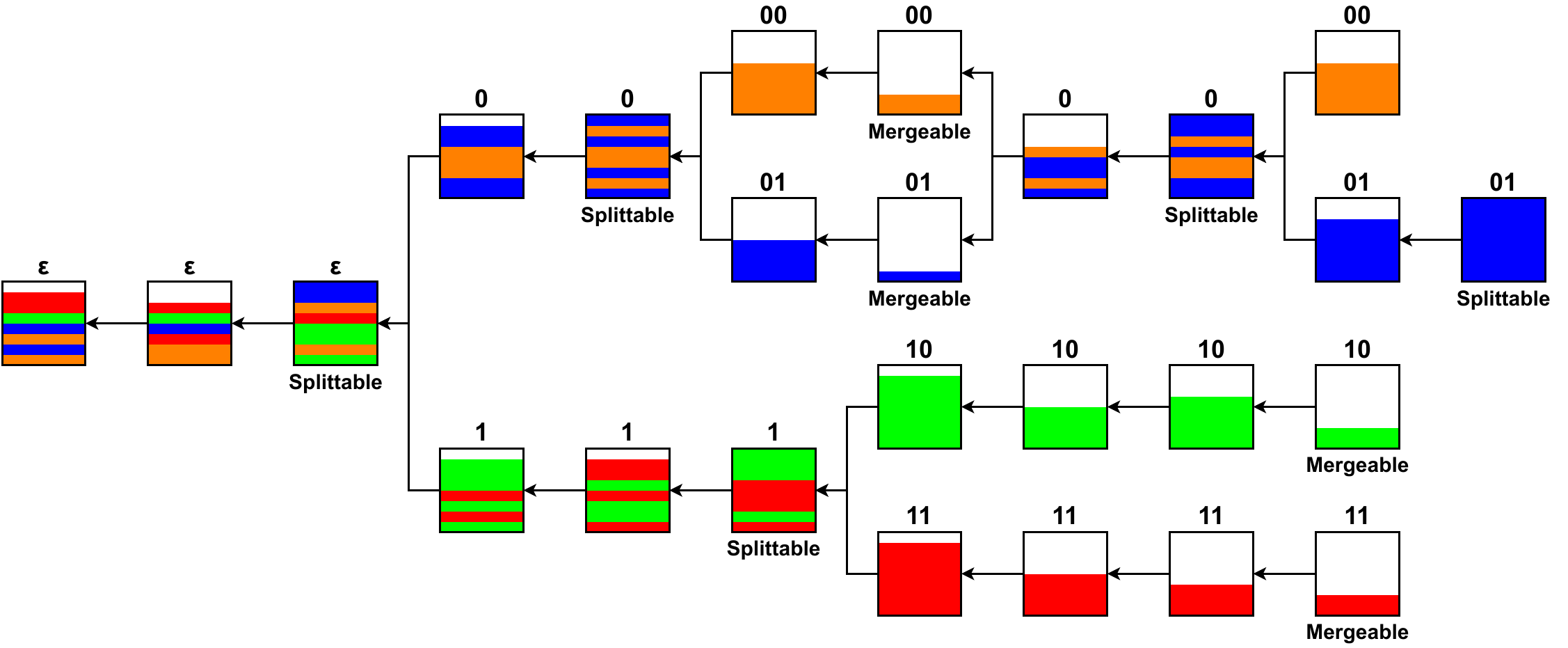}
    \caption{
        Ledger creates a split in the block DAG in times of saturation.
        Transactions are color-coded based on their prefix, which determines their position after a split.
        Proposed in~\cite{sycomore}.
    }
    \label{fig:sycomore}
\end{figure}

A solution to alleviate the issue with overlapped transactions in the blocks is to divide the transactions into separate buckets based on the prefix of their hash.
It is noteworthy that clients have no incentive to manually alter the hash, as transactions are distributed uniformly across the buckets, resulting in constant fees.
The length of the prefix, and thus the number of buckets, is dynamically adjusted based on the current number of pending proof transactions in the network (as seen in \autoref{fig:sycomore}).
Miners can select a bucket of their choice, but the miner's inclusion set for the block is strictly limited to transactions originating from that bucket.
Miners in distinct buckets work on disjoint transaction sets, eliminating duplicated effort across buckets.
Within a bucket, only the lottery winner's proofs are published; the wasted work ratio is quantified in~\autoref{fig:h4}.
Since bucket assignment is determined solely by the transaction's hash, a client attempting to target a different bucket would change the proof parameters, making it a distinct request.

This is an approach inspired by Sycomore~\cite{sycomore} that dynamically splits the blockchain into multiple parallel canonical branches according to network demand (\autoref{fig:sycomore}).
The proposed adjustment would mean that the proof and coin transaction space would be logically partitioned. By partitioning the transaction space, overlap can occur only within the same bucket, thereby enabling parallelism and significantly reducing the likelihood of conflicts between concurrently mined blocks, thus reducing wasted work (see $\mathbf{H_4}$ in \autoref{sec:eval}).

\subsection{Integrity of the Blockchain}
\label{sec:auth}

To prevent the mined solution (proof) from being stolen and used in another block, each arithmetic circuit must be slightly modified by introducing a new public parameter, which we call the \textit{integrity parameter} $\eta$.
The integrity parameter is initially calculated by hashing the header of the block being produced, which contains the previous block’s hash, along with the included coin transactions.
Next, the integrity parameter is injected along with client-provided parameters into the circuit (as shown in~\autoref{fig:lottery}) and a proof is generated.
If a lottery fails, the integrity parameter is recalculated to also include the newly produced proof to prevent miners from discarding proofs that did not yield a lottery win, producing a cryptographically linked chain of proofs.

\hyperref[alg:zokrates-source]{Algorithm 1} shows an example source code of a program in ZoKrates DSL that proves knowledge of two factors of a certain number without revealing the factors to the verifier. 
To integrate this program into our system, we added an additional public parameter \texttt{integrity} to ensure integrity (as described in \autoref{sec:auth}) along with a dummy \texttt{assert} statement to prevent the argument from being removed during optimization (both changes are highlighted in orange). Altering the value of the integrity parameter $\eta$ causes the proof verification to fail.

The role of the mask parameter is to ensure that every input of the circuit is unique and thus avoid the amortization of work, as pointed out by Kattis and Bonneau in their Proof of Necessary Work design~\cite{ponw}. 
The mask is computed as a combination of the miner's address and the number of times the particular circuit was utilized by the miner. 
See further discussion on amortization in \autoref{sec:discussion}.
In Groth16, changing any public input invalidates the entire QAP evaluation, requiring full recomputation of all witness polynomials and group exponentiations. Grinding over $\eta$ or the mask thus offers no shortcut over honest proving.

\begin{algorithm}[t]
\caption{An example of the arithmetic circuit with highlighted necessary modifications.}
\label{alg:zokrates-source}
\scriptsize
\begin{Verbatim}[commandchars=\\\{\}]
\textcolor{cb_blue}{def} main(    
    \textcolor{cb_blue}{private} \textcolor{cb_green}{field} _factor1, // factor1 + mask
    \textcolor{cb_blue}{private} \textcolor{cb_green}{field} _factor2, // factor2 + mask
    \textcolor{cb_blue}{public} \textcolor{cb_green}{field} product,
    \highlight{\textcolor{cb_blue}{public} \textcolor{cb_green}{u256} integrity} // h(curHdr)  
    \highlight{\textcolor{cb_blue}{public} \textcolor{cb_green}{u256} mask} // miner + circCnt 
) -> \textcolor{cb_green}{bool} \{
    \highlight{\textcolor{cb_blue}{assert} (integrity != \textcolor{cb_lime}{0});} // Dummy utilization
    \highlight{\textcolor{cb_green}{field} factor1 <==  _factor1 - mask } 
    \highlight{\textcolor{cb_green}{field} factor2 <==  _factor2 - mask } 
    \textcolor{cb_blue}{assert} (factor1 * factor2 == product);    
    \textcolor{cb_blue}{return true};
\}
\end{Verbatim}
\end{algorithm}

\begin{figure*}[t]
    \centering
    \includegraphics[width=0.85\linewidth]{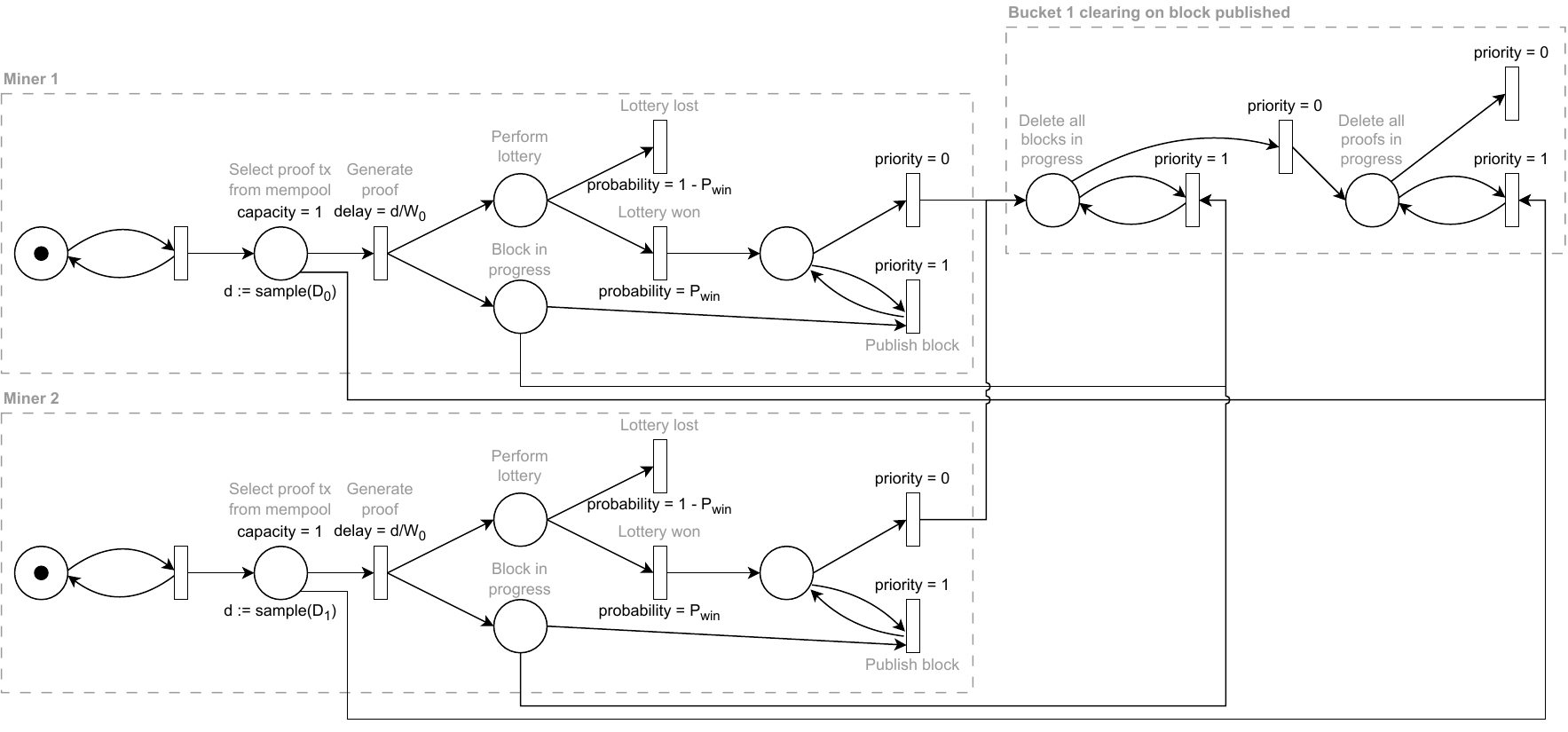}
    \vspace{-0.35cm}
    \caption{
        Stochastic Time Colored Petri Net (STCPN) modeling the consensus protocol with two miners and a single bucket.
        The network can be extended to more miners or buckets by replicating the miner or bucket component.
}
    \label{fig:petri}
\end{figure*}

\subsection{Circuit Registry Subnetwork}
\label{sec:circuit-management}
Our solution uses a subnetwork of nodes that manage the registered arithmetic circuits.
The subnetwork provides functions for decentralized registration and retrieval of arithmetic circuits, along with the corresponding proving and verification keys, circuit complexity, and the content of the original circuit source code file in the domain-specific language.
This enables clients to query the database for an existing circuit that meets their requirements, thereby avoiding the overhead of registering a new circuit and reducing unnecessary load on the subnetwork.
The circuit subnetwork consists of nodes that stake a sufficient amount of native currency.
In the event of node misbehavior (submitting an invalid SMPC share) or inactivity (failing to respond to circuit metadata requests or missing an SMPC contribution), the stake will be slashed.

A client requesting the registration of a new circuit pays a fee that is distributed among the subnetwork nodes.
The fee equals the circuit's gate count multiplied by a per-gate price constant (see~\autoref{alg:essential}).
Proof transaction fees are client-set but must exceed a complexity-proportional minimum.
The registration request contains a source code file that is compiled by subnetwork nodes, which then participate in the trusted setup by contributing randomness using SMPC.
SMPC enables nodes to generate the circuit key pair such that security is preserved as long as at least one node is honest and deletes its randomness after the ceremony.

\subsection{Extension: Proofs with Private Parameters}
Our current solution can only generate SNARKs with public parameters, as clients must publicly share the requested proof parameters in the mempool. 
While some applications can operate without private parameters and rely on the network state as-is, many others (e.g., privacy-preserving ones) require proofs that incorporate private inputs.
Therefore, we propose an approach to overcome this limitation through the witness obfuscation outsourcing technique, which is described in \autoref{sec:zk-extension} of the Appendix (due to length limitations).

\section{Evaluation}\label{sec:eval}

We modeled the lottery portion of our protocol using a Petri net (\autoref{fig:petri}) to empirically verify the following hypotheses:

\begin{itemize}
    \item[$\bullet$] $\mathbf{H_1}$: Reward distribution is commensurate with the mining power of nodes.
    \item[$\bullet$] $\mathbf{H_2}$: The complexity of selected proofs during a certain time frame does not affect the reward.
    \item[$\bullet$] $\mathbf{H_3}$: Introduction of the $\psi$ parameter (\autoref{sec:psi-parameter}) reduces wasted work at the cost of increasing centralization.
    \item[$\bullet$] $\mathbf{H_4}$: Introduction of bucketing (\autoref{sec:bucketing}) reduces wasted work.
\end{itemize}

\begin{figure}[b]
    \centering
    \includegraphics[width=0.65\linewidth]{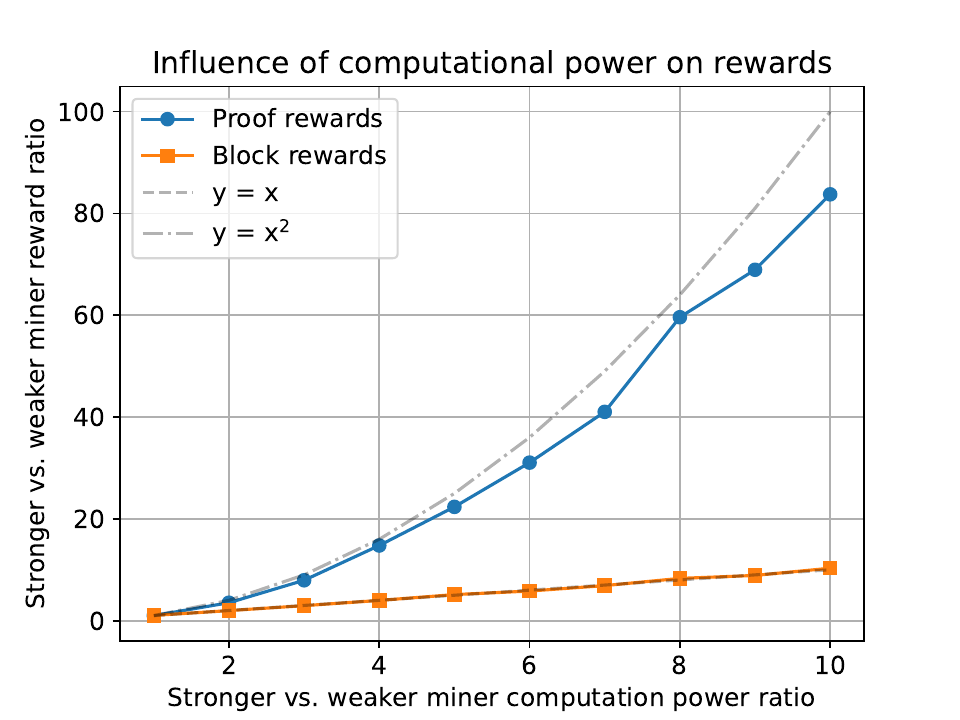}
    \caption{
        Simulation results for $\mathbf{H_1}$ run with 2 miners, where the relative computational power of a stronger miner was increasing as indicated on the $x$-axis.
    }
    \label{fig:h1}
\end{figure}

We implemented the Petri net (\autoref{fig:petri}) in Python using the \texttt{simpn} library.\footnote{\url{https://pypi.org/project/simpn/}}
We focused on the proof generation process and the resulting interactions among miners. 
Consequently, our model considered only proof transactions, as coin transaction confirmation is effectively instantaneous by comparison.
Likewise, block verification time was negligible and therefore omitted. 
We assumed that each miner had already cached the data required to generate a proof from the circuit registry subnetwork (\autoref{sec:circuit-management}).

\subsubsection{\textbf{Plain Lottery}}
\begin{figure}[t]
    \centering

    \includegraphics[width=0.65\linewidth]{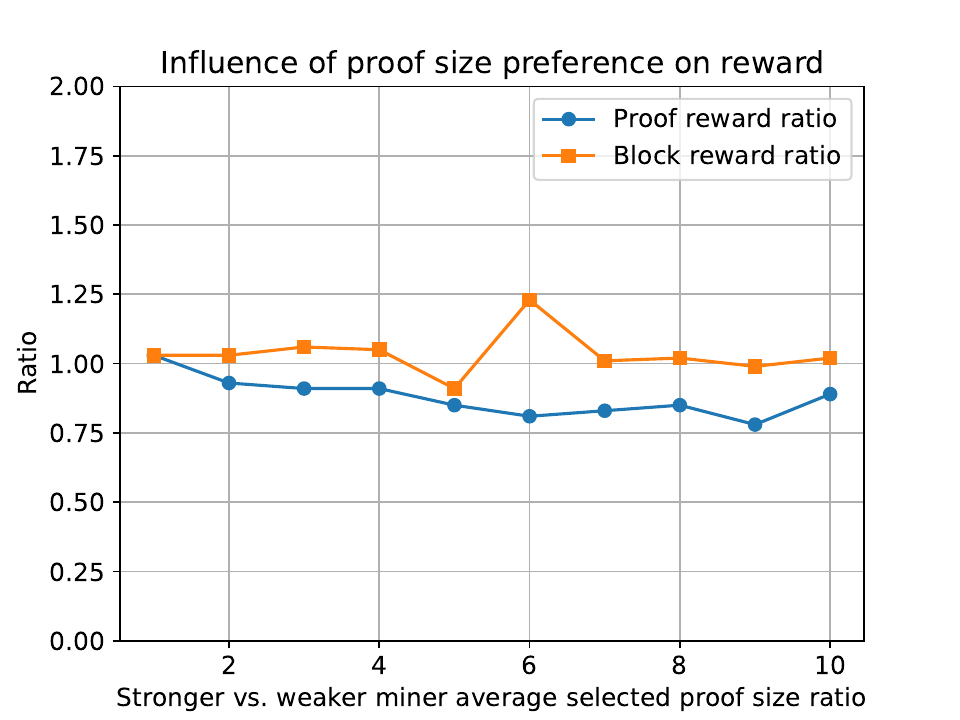}
    \includegraphics[width=0.65\linewidth]{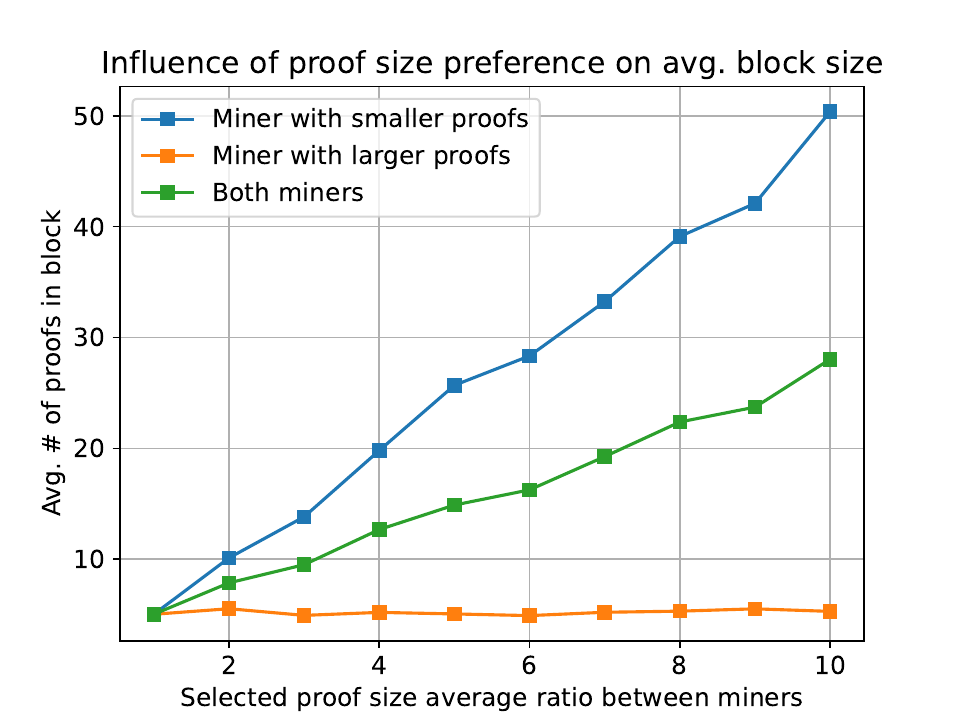}
    
    \caption{
        Results of experiments for hypothesis $\mathbf{H_2}$ regarding the preference of proof complexity.
        Simulation was run with two miners of the same computational power, but with a different preference of proof size (as indicated by the $x$-axis).
    }
    \label{fig:h2}
\end{figure}

Plain lottery in its simplest form requires hypotheses $\mathbf{H_1}$ and $\mathbf{H_2}$ to provide the fairness property.
Experiments for these hypotheses were run with two miners which differed in their computational power by a certain factor in each run, as shown in \autoref{fig:h1} and \autoref{fig:h2}.
Block rewards ratio for miners has an identity relationship with the computational power ratio, as expected.
However, the proof rewards ratio exhibits a quadratic dependence, which is caused by the stronger miner not only producing more blocks, but also each block being stronger on average compared to the weaker miner.
This quadratic trend would make the reward system unfair since increasing consensual power by a factor of two would increase proof rewards four-fold.
Thus, the contribution of proof rewards in total miner rewards should be minimized to disincentivize the creation of mining pools.
However, proof rewards cannot be removed because miners need to be motivated to continue mining the same block after an unsuccessful lottery instead of starting a new block from scratch.

\subsubsection{\textbf{Lottery Preferring Longer Chains}}
\begin{figure}[t]
    \centering

    \includegraphics[width=0.65\linewidth]{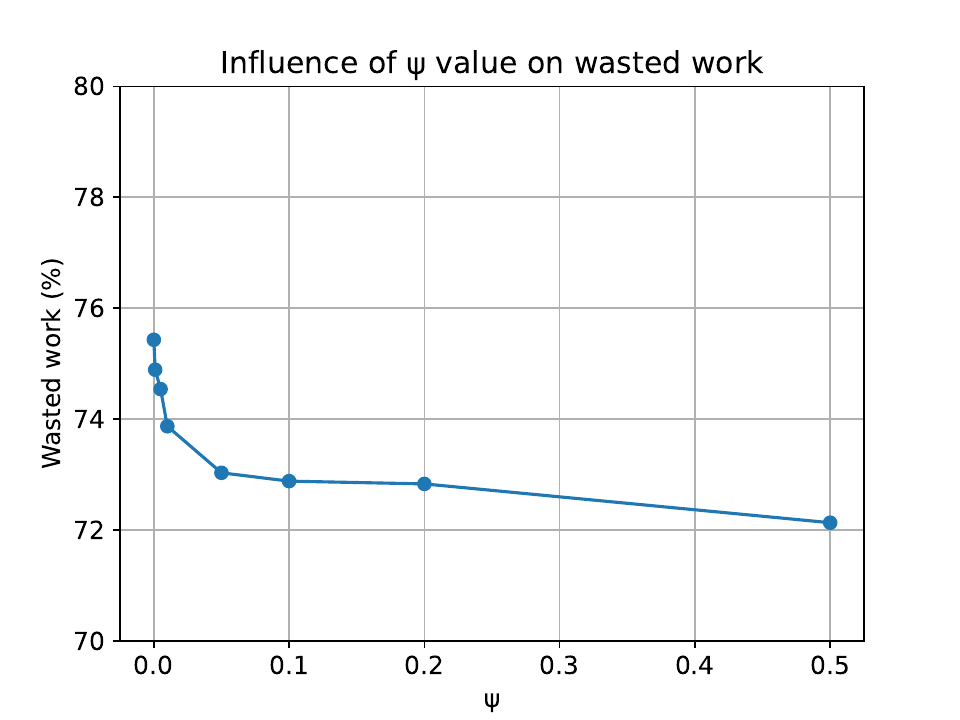}
    \includegraphics[width=0.65\linewidth]{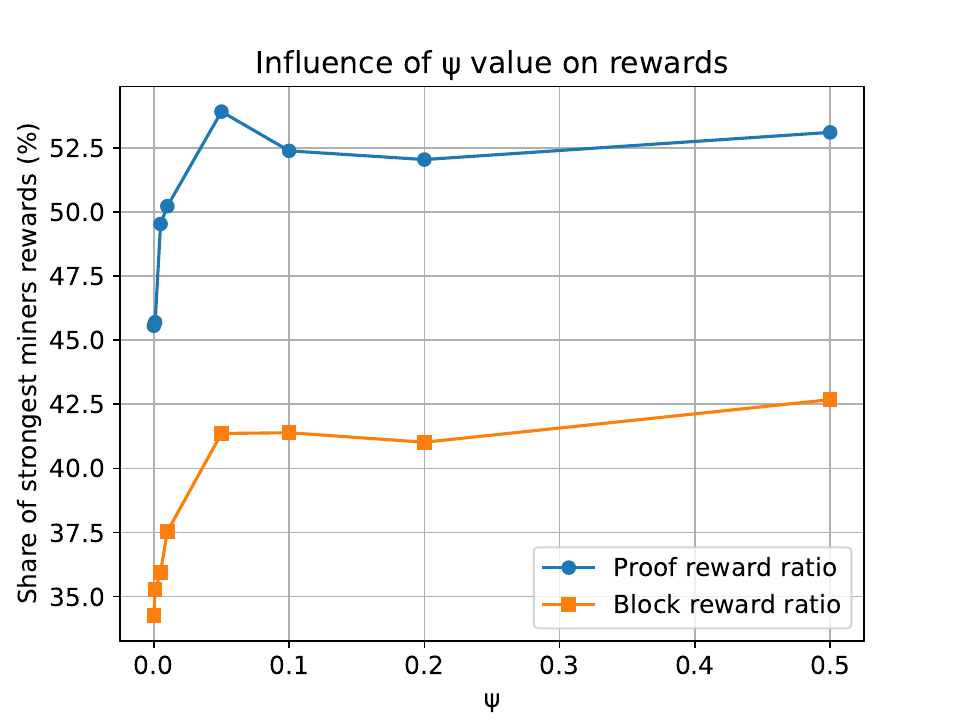}
    
    \caption{
        The experimental results of $\psi$ parameter addition ($\mathbf{H_3}$) with 5 miners with different computational power.
    }
    \label{fig:h3}
\end{figure}

In \autoref{sec:psi-parameter} we introduced the parameter $\psi$ into the lottery formula, which controls the influence of current proof chain complexity on the lottery chance, as a means to prefer longer proof chains. We anticipated that with the increasing value of this parameter the network would decrease the ratio of wasted work to all performed work at the cost of reducing fairness (hypothesis $\mathbf{H_3}$). \autoref{fig:h3} shows the evolution of wasted work and rewards distribution with regard to the increasing parameter $\psi$.

\subsubsection{\textbf{Bucketed Lottery}}
\begin{figure}[b]
    \centering
    \includegraphics[width=0.65\linewidth]{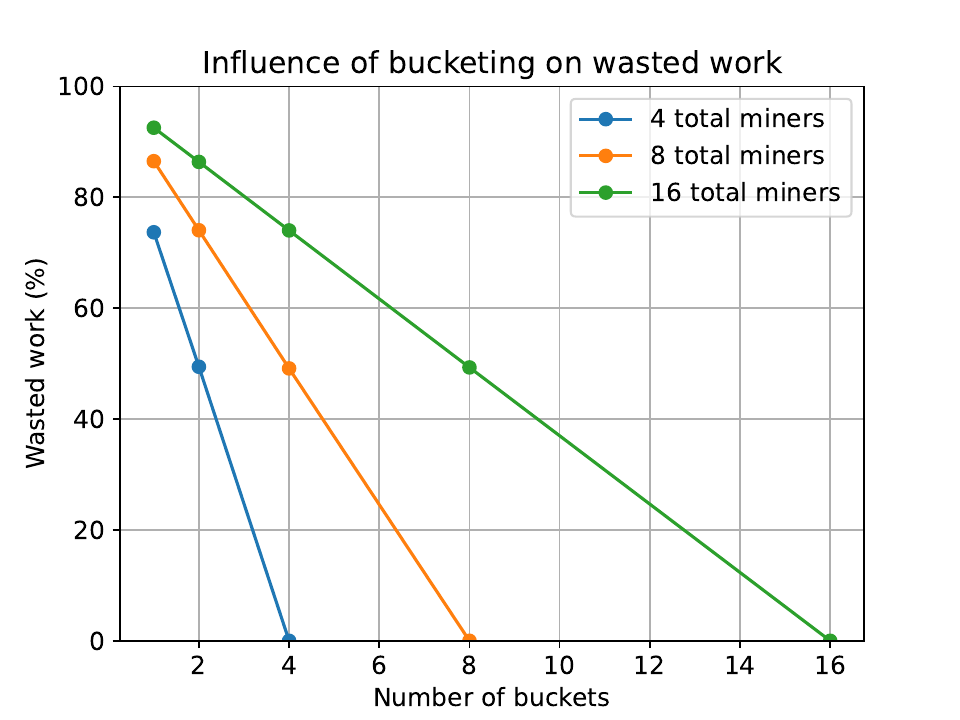}
    \caption{
        Results from simulation of $\mathbf{H_4}$.
    }
    \label{fig:h4}
\end{figure}

Hypothesis $\mathbf{H_4}$ considers the influence of the number of buckets on the amount of wasted work (\autoref{fig:h4}). It is clear from the result that a high number of buckets is beneficial; however, their number is constrained by the number of proofs that are available in the mempool.
 
\section{Discussion}\label{sec:discussion}

In this section, we examine specific aspects of our proposed solution, focusing especially on how our consensus protocol fulfills the key PoW conditions.

\subsubsection{\textbf{Time to Generate a Proof}}
\label{sec:time-to-generate}

Proof generation time in Groth16 zk-SNARKs\cite{groth2016snarks} is critical for the scalability of our consensus protocol and block time calibration.
A linear relationship between the number of constraints and proof time is essential for predictable performance and efficient resource allocation.
This linearity allows workers to handle increasingly complex circuits without exponential slowdowns, supporting the decentralized and scalable nature of the protocol.
Theoretically, the proof time \( T(n) \) for a circuit with \( n \) constraints follows $T(n) = a \cdot n + b$, where \( a \) is the per-constraint cost (e.g., polynomial evaluations) and \( b \) is the fixed overhead (e.g., constant-time cryptographic routines and I/O operations).

As confirmed by benchmarking data from~\cite{zk-bench} and the protocol comparison from~\cite{protocol-comparison}, Groth16 exhibits a linear trend in proof time with increasing constraint count.
With a precomputed trusted setup, proof generation becomes the dominant cost and scales predictably with circuit complexity.

\subsubsection{\textbf{Alternative Proving Schemes}}
In our proposal, we only explored the protocol variant that uses SNARKs with the Groth16 proving scheme.
Groth16 is a good fit due to two properties -- its constant proof size and its proving time being asymptotically linear with respect to the circuit’s constraint count (see \autoref{sec:time-to-generate}).
The ability to verify block correctness necessitates the storage of proofs on-chain.
With Groth16, the actual proof only takes up 128\,B.
Alternative proving schemes could be substituted, provided they maintain constant proof size and their proving time is predictable from the circuit's constraint count (not necessarily linear).
Light client support via recursive proof composition or aggregation is left as future work.

\subsubsection{\textbf{Puzzle Condition Analysis}}
In \autoref{sec:pow-conditions}, we outlined the essential conditions for a secure and effective PoW protocol. The following discusses how our proposed protocol addresses each of these requirements:
\begin{itemize}
    \item[$\bullet$] \textbf{Asymmetry}: 
    SNARK proof generation and verification time fits very well into this condition.
    The verification time is in the order of milliseconds and is constant with respect to the complexity of the related circuit (see \autoref{sec:time-to-generate}).

    \item[$\bullet$] \textbf{Granularity and Parametrization}: 
    Target difficulty can be adjusted to control the expected block time, allowing for fine-grained tuning of the block production rate.

    \item[$\bullet$] \textbf{Amortization-freeness}:  
    While methods for batching the verification of Groth16 proofs exist~\cite{groth16-batching}, to our knowledge, no analogous techniques have yet been developed for batching the proving process.
    Although batching methods for proof generation are being actively explored~\cite{zkp-batching}, these efforts currently focus on protocols other than Groth16.

    \item[$\bullet$] \textbf{Independence}: 
    There is no work shared between the generation of two separate proofs, and no advantage is gained for generating one after the other, even if they use the same circuit with different parameters.
    Even if there were two proof transactions in the mempool with the same circuit and parameters, their proof generation would still constitute completely independent work, since they differ in the integrity parameter.

    \item[$\bullet$] \textbf{Efficiency}: 
    The network introduces some inefficiencies, such as the use of SMPC for the trusted setup.
    However, these costs are amortized as circuits are reused with different parameters, which is a behavior encouraged by the high registration fee for new circuits.
    The actual proof computation is modified by introducing a single new parameter that ensures integrity, but this change has minimal impact on resource usage.

    \item[$\bullet$] \textbf{Unforgeability}: 
    At first glance, it may appear that a miner could create their own pending proof transactions, withhold them from the network, and begin secretly precomputing proofs.
    However, because the integrity parameter of proofs depends on the previous block’s hash, miners cannot precompute or forge proofs in advance without knowing the future chain state, preventing secret manipulation of pending proofs.
    
    \item[$\bullet$] \textbf{Freshness}: 
    As mentioned in \autoref{sec:auth}, the integrity parameter of each proof serves as a binding commitment between the proof and its containing block, ensuring that the proof cannot be reused or hijacked in another block without invalidating verification.

    \item[$\bullet$] \textbf{Deterministic}:
    The cost of a single proof is deterministic, with a linear relationship between constraint count and proving time (\autoref{sec:time-to-generate}).
    The cost of winning a block is stochastic by design, but its expected value is predictable from $\kappa$.
    
    \item[$\bullet$] \textbf{Progress-free}: 
    Each completed proof triggers an independent Bernoulli trial via the lottery (\autoref{sec:lottery}).
    Internal progress within a single proof resets upon completion - the lottery outcome depends only on the current proof's complexity.

    \item[$\bullet$] \textbf{Publicly Verifiable}: 
    Proofs are published on the blockchain for public verification, with the verifying key provided by any member of the circuit registry subnetwork (see \autoref{sec:circuit-management}).
\end{itemize}

\subsubsection{\textbf{Solution Amortization Prevention}}

Amortization undermines consensus fairness by enabling miners to generate proof batches with disproportionately less effort than single instances.
In Groth16, we distinguish between unavoidable yet negligible system-level amortization, such as precomputing FFT tables or loading the ZKP library into memory, and cryptographic amortization stemming from the mathematical properties of the proving scheme. Unlike universal schemes such as PLONK or HALO which may support cross-circuit amortization, Groth16 proving keys encode circuit-specific structure from the trusted setup, so keys for logically related circuits (e.g., f, g, and f $\wedge$ g) share no algebraic structure and proving one yields no advantage toward proving another. While limiting miners to a single proof per circuit prevents amortization, it significantly throttles throughput. A more robust solution utilizes Pedersen hashing to pseudorandomly alter all parameter values between proofs. This effectively prevents amortization, though it restricts the set of available operations within the modified circuit~\cite{ponw}.

\section{Related Work}\label{sec:related}

This section reviews existing literature in two domains: Proof of Useful Work (PoUW) protocols and marketplaces for SNARK proof generation.

\paragraph{\textbf{Proof of Useful Work}}

Primecoin~\cite{primecoin}, introduced in 2013 pioneered practical PoUW by having miners search for prime number chains, contributing to number theory.
Ball et al.~\cite{pouw} formalized PoUW using hard problems such as orthogonal vectors, addressing the usefulness through randomized challenges of prior block hashes.
Later works applied NP-hard problems, such as combinatorial optimization in Ofelimos~\cite{ofelimos} and minimal dominating sets in Chrisimos~\cite{chrisimos}.
Other works~\cite{sym14091831,PUPoW} have focused on creating frameworks for solving general real-life optimization problems.
Other approaches use matrix multiplication~\cite{multiplication} or machine learning, such as distributed deep learning in Coin.AI~\cite{coinai} and scalable ML training in PoGO~\cite{pogo} through quantized gradients and Merkle proofs.
Generalized frameworks encompass PUPoW~\cite{PUPoW} for practical blockchains, Proofware~\cite{proofware} for dApps, COCP~\cite{cocp}, and crowdsourcing paradigms~\cite{crowdsourcing} to integrate general real-life tasks into the mining process.
Nevertheless, none of these works satisfy all the required properties of PoW (see \autoref{sec:problem}). In particular, freshness is the most critical property, as its absence allows adversaries to appropriate recently computed solutions.

\paragraph{\textbf{Proof Generation Marketplaces}}

The computational demands of the generation of the zk-SNARK proof have spurred marketplaces, divided into consensus-integrated and standalone models.
Consensus-integrated systems~\cite{ponw} embed proof in blockchain consensus using PoUW SNARKs for state verification, although limited to fixed purpose-specific circuits without arbitrary proof support.
In particular, Mina protocol~\cite{mina} employs recursive SNARKs to maintain a constant state (as well as a constant size), but restricts recursion to block validation.
Its Snarketplace~\cite{mina-snarketplace} enables block producers to buy proofs from SNARK workers, operating above the consensus layer.
In contrast, our approach integrates PoUW directly at the consensus layer, requiring miners to fulfill SNARK proof generation requests for valid block creation.
Standalone networks such as =nil; Foundation Proof Market~\cite{nil-proof-market} offer a decentralized exchange with SNARK order books, and RISC Zero's Bonsai~\cite{risc-zero-bonsai} as a zkVM-based coprocessor for arbitrary programs in standard languages for dApps.

\section{Conclusion}\label{sec:conclusion}
Due to PoW's high energy use and environmental impact, alternatives with their specific trade-offs have emerged, where PoUW offers a promising yet challenging approach to maintaining PoW's security without compromising its core properties.
Additionally, zk-SNARKs are critical for scalability and privacy in many block\-chain networks, but their generation is too resource-intensive for many users. 
Several zk-SNARK marketplaces emerged to offload their generation.

We design the PoUW protocol where SNARK proof generation constitutes the work in the PoUW consensus mechanism, providing chain security benefits of PoW while producing SNARK proofs valuable in broader cryptographic applications as a byproduct.
To the best of our knowledge, our solution is the first one that meets all the necessary properties of PoW in contrast to prior work. 
 
\bibliographystyle{IEEEtran}
\bibliography{main}

\appendix

\subsection{Block Production and Verification Pseudocode}
\autoref{alg:essential} shows pseudocode of the most important functions for block generation and verification of a block as well as registration of a proof within the circuit registration subnetwork.
\vspace{10pt}

\begin{algorithm}[H]
\caption{Pseudocode of essential functions.}
\label{alg:essential}
\begin{algorithmic}
\scriptsize
\Function{produceBlock}{$\kappa$}
    \State blkHdr $\gets$ \Call{createBlkHdr()}{} \Comment{\textcolor{gray}{Timestamp, etc.}}
    \State coinTxs $\gets$ \{\Call{createCoinbaseTx()}{}\} $\cup$ mempool.selectValidCoinTxs()
\State proofTxs $\gets$ \{\}
    \State blkHash $\gets$ \Call{H}{blkHdr $\|$ coinTxs $\|$ proofTxs}
    \vspace{0.3em}
    \Repeat \Comment{\textcolor{gray}{Build the proof chain}}
        \State proofTx $\gets$ \Call{mempool.selectValidProofTx()}{}
        \State proofTx.proof $\gets$ \Call{zkLib.genProof}{proofTx, blkHash}
        \State proofTxs $\gets$ proofTxs $\cup$ \{proofTx\}
        \State blkHash $\gets$ \Call{H}{blkHdr $\|$ coinTxs $\|$ proofTxs}
    \Until{blkHash $<$ $\kappa$ * proofTx.complexity} \Comment{\textcolor{gray}{Lottery-weighted diffic. threshold}}
    \vspace{0.3em}
    \State \Return{blkHdr, coinTxs, proofTxs}
\EndFunction
\vspace{1em}
\Function{verifyBlock}{blk}
  \State assert(isBlkHdrValid(blk.hdr)) \Comment{\textcolor{gray}{Timestamp, etc.}}
  \State assert(areCoinTxsValid(blk.coinTxs))
\vspace{0.3em}
  \State checkedProofTxs $\gets$ \{\}
  \vspace{0.3em}
  \State $\color{gray}{\triangleright}$\,\textcolor{gray}{Iteratively build and verify proof chain}
  \For{proofTx \textbf{in} blk.proofTxs}
      \State iterHash $\gets$ \Call{H}{blk.hdr $\|$ coinTxs $\|$ checkedProofTxs}
      \State assert(zkLib.verifyProof(proofTx, iterHash))
      \State checkedProofTxs $\gets$ checkedProofTxs $\cup$ \{proofTx\}
  \EndFor
\EndFunction
\vspace{1em}
\Function{genProof}{proofTx, integrity}
  \State hash $\gets$ proofTx.circuitHash
  \State circuit, complexity, provingKey $\gets$ registry.getCircuit(hash)
  \State args $\gets$ proofTx.parameters $\|$ integrity
  \State proof $\gets$ zkLib.generateProof(circuit, provingKey, args)
  \vspace{0.3em}
  \State \Return proof
\EndFunction
\vspace{1em}
\Function{registerCircuit}{sourceCode, burnTxId}
  \State compiledCircuit $\gets$ zkLib.compile(sourceCode)
  \State complexity $\gets$ zkLib.getConstraintCount(compiledCircuit)
  \State circuitHash $\gets$ H(compiledCircuit)
  \vspace{0.3em}
  \If{registry.doesCircuitExist(circuitHash)}
    \State \Return{circuitHash} \Comment{Reject duplicate circuit registration}
  \EndIf
  \vspace{0.3em}
  \State $\color{gray}{\triangleright}$\,\textcolor{gray}{Perform SMPC for trusted setup and register circuit}
  \State feeTx $\gets$ blockchain.getTxById(burnTxId)
  \State assert(isValid(feeTx))
  \State assert(feeTx.amount $>=$ complexity * pricePerGate)
  \vspace{0.3em}
  \State provingKey, verifKey $\gets$ registry.performSMPC(compiledCircuit)
  \State registry.store(circuitHash, complexity, provingKey, verifKey)
  \State registry.distributeRewards(feeTx.amount) \Comment{\textcolor{gray}{To SMPC contributing nodes}}
  \vspace{0.3em}
  \State \Return circuitHash
\EndFunction

\end{algorithmic}
\end{algorithm}
 
\newpage
\subsection{Zk-SNARK Puzzle Extension}
\label{sec:zk-extension}

The current zk-SNARK marketplace proposal requires all inputs, including the private witness, to be publicly submitted to the block\-chain, breaking the zero-knowledge property essential to zk-SNARKs.
Supporting private inputs in outsourced proof generation is critical to maintaining privacy while leveraging external workers for computation.
Several techniques have been explored, ranging from basic approaches to advanced cryptographic methods, evaluated based on key parameters: \textbf{confidentiality} (ensuring private inputs remain hidden from workers), \textbf{open-proving} (allowing any participant to act as a worker without restrictions), \textbf{non-interactivity} (requiring no communication between client and worker), and \textbf{feasibility} (practicality in terms of computational and integration complexity).

\paragraph{\textbf{Naive Approach}}
A basic approach is to directly share private inputs with the external workers for zk-SNARK proof generation.
The client submits a proof request on-chain and sends the inputs to any worker via a secure channel.
If multiple workers are considered, inputs must be sent to each, increasing communication overhead.
This method fully compromises confidentiality, as workers gain complete access to sensitive data, violating the zero-knowledge property.

\paragraph{\textbf{Fully-Fledged Options}}
To overcome the limitations of a naive approach, we outlined several options, drawing from established methods in secure computing and cryptography:

\begin{itemize}
    \item[$\bullet$] \textbf{Multiparty Computation (MPC)}: 
The client splits the witness using a secret-sharing scheme (e.g., Shamir’s~\cite{shamir1979secret}) and distributes shares to multiple workers. A secure MPC protocol~\cite{yao1986mpc} enables joint proof generation without revealing the full input. The final proof is aggregated and submitted on-chain.
    MPC ensures confidentiality and supports open-proving, but the overhead increases with the circuit size and the worker count. Fewer nodes increase the risk of Sybil risks. Integration is complex due to consensus modifications.
    
    \item[$\bullet$] \textbf{Fully Homomorphic Encryption (FHE)}: 
The client encrypts the witness using FHE~\cite{gentry2009fhe} and includes it in the proof request. Workers compute the proof homomorphically and return an encrypted result without seeing the inputs.
    FHE provides strong privacy and non-interactivity, enabling open-proving. However, it remains highly inefficient for large circuits.
No end-to-end zk-SNARK FHE solution is practical as of now.

    \item[$\bullet$] \textbf{Trusted Execution Environments (TEEs)}:
A TEE-enabled worker decrypts the witness inside a secure enclave (e.g., Intel SGX~\cite{costan2016sgx}) and generates the proof, submitting it with a hardware attestation.
    TEEs offer confidentiality via hardware isolation and enable non-interactive proofs, but they restrict proving to specific hardware, introduce the necessity for vendor trust, and add complexity due to attestation and environment adaptation.

    \item[$\bullet$] \textbf{Witness Obfuscating Outsourcing (WOO)}: 
    The client blinds each private input \( s_{\text{in}} \) with a random value \( r_{\text{in}} \), producing \( s'_{\text{in}} = s_{\text{in}} + r_{\text{in}} \) \cite{nakamura2023outsourcing}. The obfuscated witness is included in the proof request, allowing the worker to generate the proof without accessing the original inputs.
    WOO provides strong confidentiality and enables open-proving without special hardware. It is non-interactive and integrates well with existing zk-SNARK tools, adding minimal overhead.
    The main drawback is the need to adapt the circuits to handle obfuscated inputs, requiring a light preprocessing step on the client side. This step remains compatible with standard zk-SNARK frameworks.
\end{itemize}

\noindent
WOO stands out for its balance of confidentiality, open-proving, non-interactivity, and feasibility, making it the preferred solution for maintaining privacy in outsourced zk-SNARK proof generation within our decentralized marketplace.

\subsection{Encrypting Private Inputs}
To protect the confidentiality of the private inputs, the client uses additive masking. For each private input \( s_{\text{in}} \), a random value \( r_{\text{in}} \) is sampled uniformly from a large finite field. The obfuscated input is calculated as:
\begin{eqnarray}
\hat{s}_{\text{in}} = s_{\text{in}} + r_{\text{in}} \label{eq:masking}.
\end{eqnarray}
These masked inputs form the obfuscated witness included in the proof transaction submitted to the blockchain mempool. The client keeps \( r_{\text{in}} \) secret, ensuring the obfuscated values reveal no information about the original inputs. This provides information-theoretic security, as \( \hat{s}_{\text{in}} \) appears uniformly random to any observer without \( r_{\text{in}} \). The process is computationally lightweight, making it highly suitable for integration into our PoUW protocol.

\begin{figure}[t]
\centering
\includegraphics[width=0.8\linewidth]{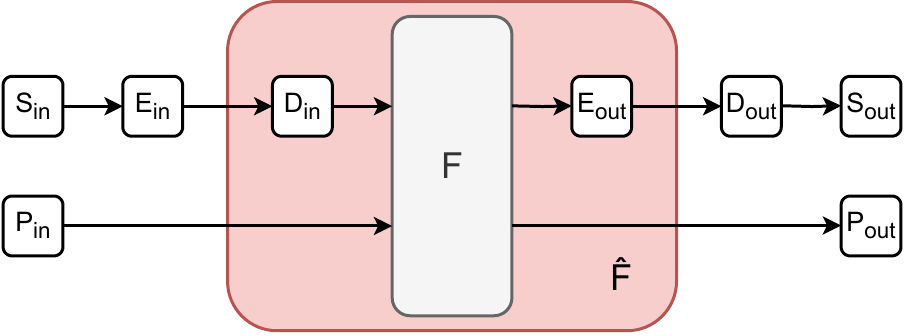}
\caption{Structure of the transformed circuit \( \hat{F} \), where encrypted input \( \hat{s}_{\text{in}} \) passes through the input decryption function \( D_{\text{in}} \), and the output \( \hat{s}_{\text{out}} \) is later decrypted using \( D_{\text{out}} \). Adapted from~\cite{nakamura2023outsourcing}.}
\label{fig:woo_architecture}
\end{figure}

\subsection{Circuit Compilation and Masking Parameters}
The masking parameters \( r_{\text{in}} \), controlling how random values are applied, are provided as function arguments during circuit compilation. Using frameworks like Zokrates~\cite{zokrates} or Circom~\cite{circom}, the high-level computation is converted into an R1CS form suitable for zk-SNARK proving.
To support masked inputs, the circuit includes an input decryption function \( D_{\text{in}} \), which recovers the original value internally using:
\begin{eqnarray}
s_{\text{in}} = \hat{s}_{\text{in}} - r_{\text{in}}. \label{eq:unmasking}
\end{eqnarray}
Here, \( r_{\text{in}} \) is hard-coded during compilation to ensure the logic cannot be tampered with. The resulting circuit \( \hat{F} \), shown in \autoref{fig:woo_architecture}, preserves the original computation but operates entirely on obfuscated inputs, allowing private and verifiable proofs.

\subsection{Trusted Setup and Proving Key}
After compiling the circuit, a trusted setup, typically via MPC among circuit registry nodes, generates the proving and verification keys.
The client contributes initial entropy, embedding masking parameters into the proving key while keeping the R1CS and masks private.
Only the proving key and obfuscated witness are sent to the worker, keeping inputs and the circuit hidden.
The verification key is published on-chain, allowing standard proof verification without changes to the consensus protocol.

\subsection{Proof Workflow}
The WOO-based proof process is divided into two stages:

\paragraph{\textbf{Client’s Workflow}}
\begin{enumerate}
    \item \textbf{Write WOO Circuit}: Design a circuit with input decryption logic \( D_{\text{in}} \) using Circom or Zokrates.
    \item \textbf{Generate Randomness}: Sample random \( r_{\text{in}} \in \mathbb{F} \) for each private input \( s_{\text{in}} \).
    \item \textbf{Compile Circuit}: Compile into an R1CS circuit with \( r_{\text{in}} \) hard-coded for use with \( \hat{s}_{\text{in}} \).
    \item \textbf{Delegate Trusted Setup}: Launch an MPC-based trusted setup, contributing initial entropy.
    \item \textbf{Generate Obfuscated Witness}: Apply additive masking to produce \( \hat{s}_{\text{in}} \).
    \item \textbf{Request Proof}: Submit a proof transaction containing \( \hat{s}_{\text{in}} \), keeping \( r_{\text{in}} \) secret.
\end{enumerate}

\paragraph{\textbf{Worker’s Workflow}}
\begin{enumerate}
    \item \textbf{Select Proof Request}: Retrieve a proof transaction from the mempool with the obfuscated witness and a reference to circuit \( \hat{F} \).
    \item \textbf{Generate zk-SNARK Proof}: Using the proving key, compute a standard zk-SNARK proof (e.g., via Groth16 \cite{groth2016snarks}) from the obfuscated witness.
    \item \textbf{Submit Block}: Broadcast a block containing the proof, meeting PoUW consensus difficulty along with other transactions.
\end{enumerate}

\subsection{Wrapping Metadata for Integrity and Authentication}
To enforce contextual proof binding while preserving zero-knowledge, the system wraps the original zk-SNARK proof inside an outer circuit. The inner proof is a private input, while the outer circuit includes public inputs like the previous block hash and miner address. It verifies the inner proof without exposing its contents.

Only the outer proof is published on-chain, ensuring confidentiality while binding the computation to specific metadata and blockchain state. During block validation, the network verifies the outer proof and its references, securing consensus without revealing sensitive data.

\begin{figure}[t]
\centering
\includegraphics[width=0.9\linewidth]{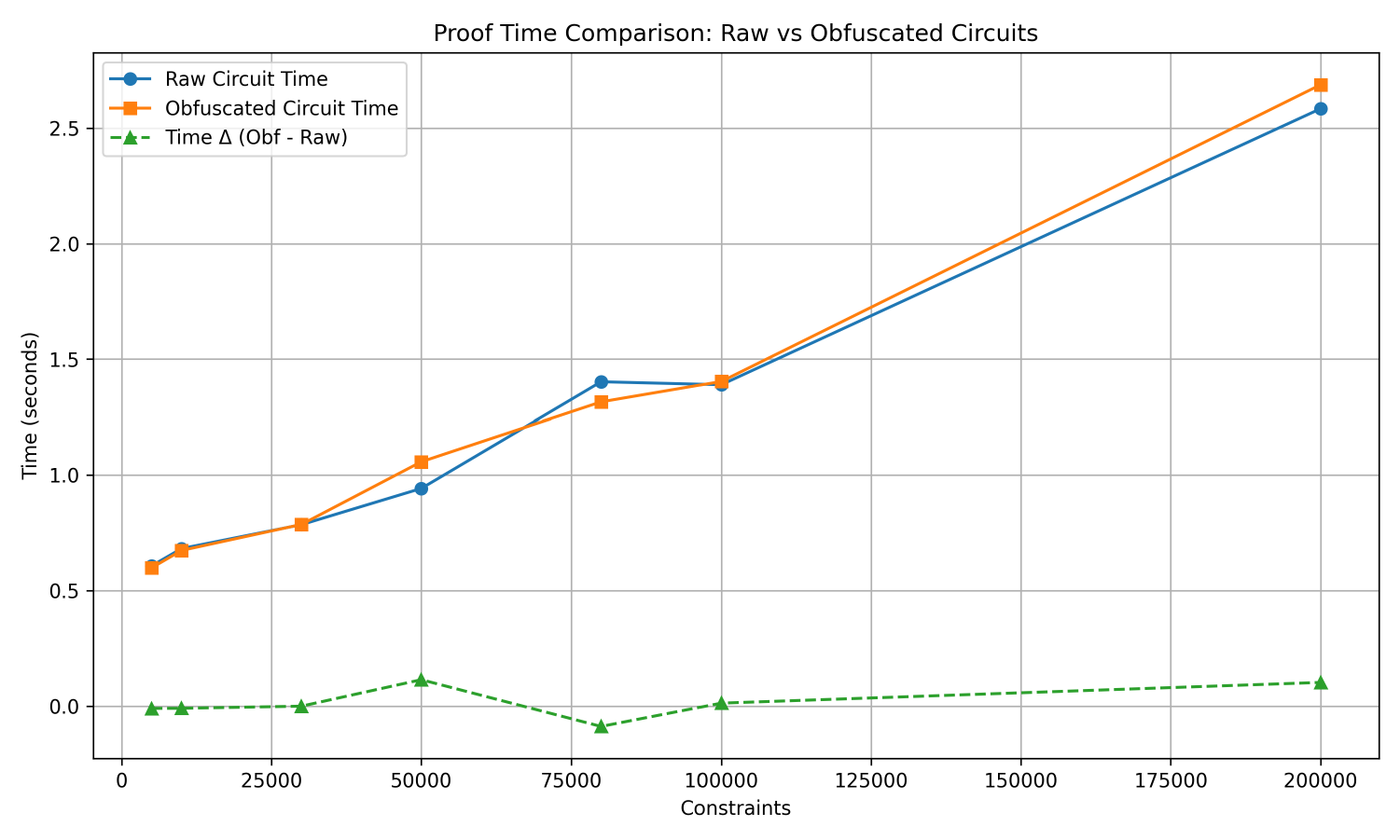}
\caption{Proof time comparison for raw vs. obfuscated Circom circuits. The green curve shows the difference between obfuscated and raw proofs (\( \Delta = T_{\text{obf}} - T_{\text{raw}} \)), which remains low across all constraint sizes. Benchmarked using Circom and snarkjs on an AMD Ryzen 5 7600X with 32\,GB RAM.}
\vspace{0.3cm}
\label{fig:mpc_vs_woo}
\end{figure}

\paragraph{\textbf{Performance Evaluation}}
To assess the impact of Witness Obfuscating Outsourcing (WOO) on proof generation time, we benchmarked obfuscated and raw circuits using Circom and Groth16, across constraint sizes from 10,000 to 200,000 under identical conditions.
As shown in \autoref{fig:mpc_vs_woo}, WOO introduces a small, consistent overhead due to embedded decryption logic. This added latency remains below 0.1 seconds, even for large circuits, and does not scale significantly with circuit size.

Overall, WOO maintains high efficiency and scalability, making it well-suited for real-time, privacy-preserving proof generation in the PoUW marketplace.

\paragraph{\textbf{Conclusion}}

The Witness Obfuscating Outsourcing (WOO) mechanism strengthens privacy in zk-SNARK proof generation by hiding private inputs from external provers using additive masking and embedded decryption logic. This enables open-proving without exposing confidential data, even in decentralized settings.

Compared to Nakamura et al.~\cite{nakamura2023outsourcing}, who rely on a trusted third party (TTP) to transform the circuit and generate the common reference string (CRS), WOO eliminates the TTP by leveraging an MPC-based setup with client-initiated entropy.
Furthermore, WOO keeps both the circuit (R1CS) and inputs hidden from workers, sharing only the proving key and obfuscated witness, which further reduces the trust assumptions and improves decentralization.
However, this enhanced privacy reduces the circuit reusability. Since obfuscation embeds client-specific randomness, circuits must be generated individually per client or transaction, increasing setup effort and reducing generality.

In summary, WOO improves confidentiality and trust minimization compared to prior approaches, but introduces a trade-off between privacy and flexibility in decentralized zk-SNARK outsourcing.
 
\end{document}